\documentclass[fleqn,usenatbib]{mnras}

\usepackage{newtxtext,newtxmath}
\usepackage{diagbox}
\usepackage{tabularx}
\usepackage{geometry}
\usepackage{rotating} 
\usepackage{cleveref}

\usepackage[T1]{fontenc}

\DeclareRobustCommand{\VAN}[3]{#2}
\let\VANthebibliography\thebibliography
\def\thebibliography{\DeclareRobustCommand{\VAN}[3]{##3}\VANthebibliography}

\usepackage{graphicx}	
\usepackage{amsmath}	

\title[ImgRes]{Image Restoration with Point Spread Function Regularization and Active Learning}

\author[Peng Jia et al.]{
Peng Jia,$^{1,2,5}$\thanks{robinmartin20@gmail.com}
Jiameng Lv,$^{1}$\thanks{lvjiameng1224@link.tyut.edu.cn}
Runyu Ning,$^{1}$
Yu Song,$^{1}$
Nan Li,$^{3}$\thanks{nan.li@nao.cas.cn}
Kaifan Ji,$^{4}$\thanks{jkf@ynao.ac.cn}
Chenzhou Cui, $^{3}$\thanks{ccz@bao.ac.cn} 
Shanshan Li $^{3}$\\
$^{1}$College of Electronic Information and Optical Engineering, Taiyuan, 030024, China\\
$^{2}$Peng Cheng Lab, Shenzhen, 518066, China\\
$^{3}$National Astronomical Observatories, Beijing, 100101,China\\
$^{4}$Yunnan Observatories, Kunming, Yunnan, ,China\\
$^{5}$Department of Physics, Durham University, Durham, DH1 3LE, UK
}

\date{Accepted XXX. Received YYY; in original form ZZZ}

\pubyear{2015}

\begin{document}
\label{firstpage}
\pagerange{\pageref{firstpage}--\pageref{lastpage}}
\maketitle


\begin{abstract}
Large-scale astronomical surveys can capture numerous images of celestial objects, including galaxies and nebulae. Analysing and processing these images can reveal intricate internal structures of these objects, allowing researchers to conduct comprehensive studies on their morphology, evolution, and physical properties. However, varying noise levels and point spread functions can hamper the accuracy and efficiency of information extraction from these images. To mitigate these effects, we propose a novel image restoration algorithm that connects a deep learning-based restoration algorithm with a high-fidelity telescope simulator. During the training stage, the simulator generates images with different levels of blur and noise to train the neural network based on the quality of restored images. After training, the neural network can directly restore images obtained by the telescope, as represented by the simulator. We have tested the algorithm using real and simulated observation data and have found that it effectively enhances fine structures in blurry images and increases the quality of observation images. This algorithm can be applied to large-scale sky survey data, such as data obtained by LSST, Euclid, and CSST, to further improve the accuracy and efficiency of information extraction, promoting advances in the field of astronomical research.
\end{abstract}




\section{Introduction}
\label{Sec:1}
Sky survey projects obtain numerous observational images containing celestial objects with extended structures, including nebulae and galaxies, which are of utmost interest and importance for astronomical research. Scientists select these targets for in-depth analysis. However, aberrations and noises in astronomical observations can degrade image quality, resulting in blurred and distorted images of celestial objects. These factors significantly affect the precision of scientific information extracted from such images. Thus, it is crucial to develop and utilize image restoration algorithms to improve image quality and enable further scientific exploration. Recent developments in deep neural networks have led to the emergence of several new image restoration algorithms, which take advantage of the powerful representation capabilities of deep neural networks and have demonstrated impressive results. Two primary types of image restoration algorithms have been proposed based on different representation strategies: algorithms based on the properties of astronomical images and those based on the properties of the degradation process, such as the point spread function (PSF) of telescopes and the characteristics of noise. These algorithms can enhance the quality of images obtained from astronomical observations.\\

Image restoration algorithms based on image properties are often developed with generative neural networks, such as the VAE \citep{kingma2013auto}, the U-Net \citep{ronneberger2015u}, or the GAN \citep{goodfellow2020generative}. These neural networks are trained on a large number of high signal-to-noise ratio (SNR) and high spatial resolution real observation images, from which they learn features. The learned features are then used to restore low SNR or low spatial resolution images. Properly selected training data can lead to effective results for galaxy images \citep{schawinski2017generative,arcelin2021deblending,jia2021data,gan2021seeinggan,li2022bsc} or solar images \citep{jia2019solar}. However, manual interventions are often required to obtain appropriate training data, and overfitting or training on improper data can result in the generation of fake structures by the neural network \citep{jia2021data}.\\

Image restoration algorithms based on the properties of PSFs aim to learn the generalized inverse function of PSFs. The method first constructs a PSF model using wavefront decomposition \citep{fetick2019physics,beltramo2019prime,fusco2020reconstruction,jia2020psf} or PSF basis \citep{jia2017blind,sun2020improving}, and then uses this model as prior knowledge to restore images through supervised or unsupervised learning. In unsupervised learning algorithms, the PSFs are decomposed into parameters, and the image restoration algorithm fits these parameters through training \citep{qi2014estimation,gao2017parametric,sureau2020deep}. However, these algorithms also face the parameter-tuning problem encountered by classical image restoration algorithms. In supervised learning algorithms, PSFs are used to generate blurred images as the training set \citep{jia2020psf,li2022galaxy,wang2022neural}. However, two issues limit the application of supervised algorithms. First, PSFs are necessary as prior information for supervised learning algorithms. Although some methods have been proposed to extract star images as PSF templates \citep{terry2022airopa}, obtaining appropriate PSFs is complicated due to the presence of sky background noise or read-out noise. Furthermore, obtaining star images from extended target images (such as galaxies or nebulae) is challenging \citep{long2019point}. Second, the neural network's generalization ability is dependent on the training set. Because PSFs can change significantly, it is difficult to obtain PSFs and develop datasets that can reflect different states of real observation data. Improper PSF datasets may introduce epistemic uncertainty, which can limit the performance of trained image restoration algorithms \citep{hullermeier2021aleatoric}.\\

Therefore, we present a novel framework for processing images from large-scale astronomical sky surveys. Our framework integrates a high-fidelity simulator of a specific telescope and a deep neural network-based image restoration algorithm with the active learning strategy. The simulator is customized for the telescope used in the sky survey and can generate simulated images with various PSFs and noise levels typically encountered in real observations. The image restoration algorithm restores these images, and we evaluate the quality of the restored images by computing the mean square error (MSE) between the original and restored images. The active learning strategy governs the simulator to generate more images with PSFs or noise levels that the image restoration neural network could not restore well, thereby further training the neural network. This approach enables us to obtain an effective neural network for image restoration. We discuss our framework in detail in Section~\ref{Sec:2} and demonstrate its ability to restore simulated and real observation images with varying levels of blur in Section ~\ref{Sec:3}. Finally, we draw our conclusions and outline future work in Section~\ref{Sec:4}.\\

\section{The Framework}
\label{Sec:2}
\begin{figure}
	\centering
	\includegraphics[width=0.45\textwidth]{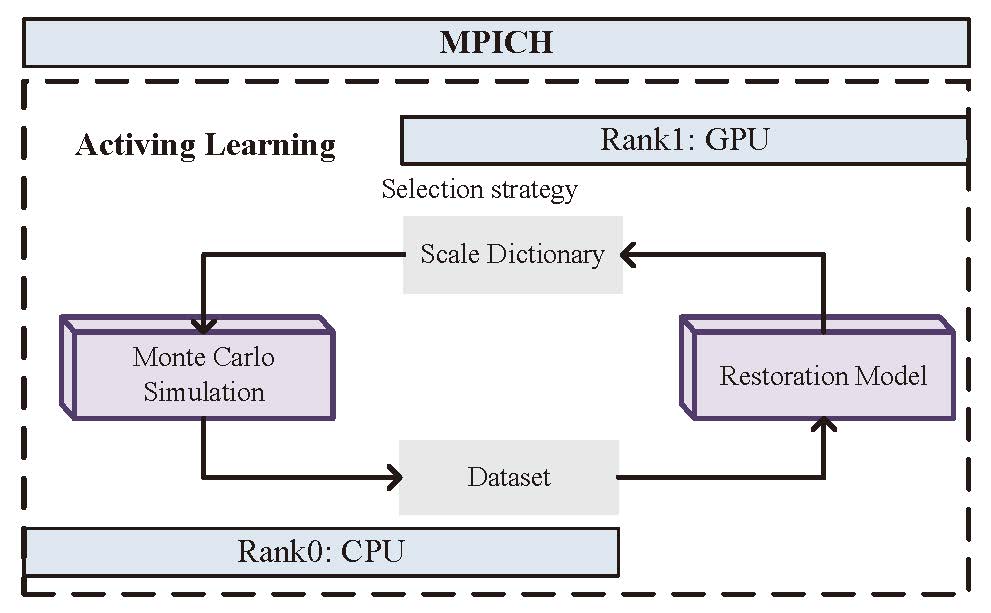}
	\caption{The image restoration framework is illustrated in the diagram, which is composed of three main parts: the Monte Carlo simulation part, the image restoration part, and the parameter selection part. In the Monte Carlo simulation part, PSFs are generated under different observation conditions, and sample images are convolved to generate blurred images as training data. The image restoration part restores these blurred images. Meanwhile, the parameter selection part evaluates the quality of the restored images and generates parameter dictionaries for the data generation part. To speed up the training process, we employ the MPI technology to run the data generation and image restoration parts in separate processors or computers.}
	\label{fig1}%
	\end{figure}

Large-scale astronomical surveys produce a vast number of blurred images with varying quality, making it impractical to manually process them one by one using human intervention-based algorithms. Therefore, an image restoration framework capable of producing stable results for images captured by the sky survey project is necessary. This paper proposes a framework that addresses this need and is illustrated in Figure~\ref{fig1}. The framework consists of three parts: the Monte Carlo simulation part, the image restoration part, and the parameter selection part. To expedite the training process of the framework, we propose using parallel computing technology to run the data generation and image restoration parts on different computers or processors. We will discuss details of our framework in the following subsections.\\

\subsection{The Monte Carlo Simulation Part}
\label{subsec21}
The Monte Carlo simulation part generates point spread functions (PSFs) of the telescope for various observation conditions and convolves sample images to create blurred images as the training data. The imaging process of an optical telescope can be modeled with the equation~\ref{eq1}: 
\begin{equation}
\label{eq0}
\begin{aligned}
	Img(x,y) = [Obj(x,y)*PSF(x,y)]_{pixel(x,y)}+Noise(x,y) 
	\end{aligned}
	\end{equation} 
where $Obj(x,y)$ and $Img(x,y)$ are original and observed images. $PSF(x,y)$ is the point-spread function (PSF) of the telescope. $[]_{pixel(x,y)}$ stands for the pixel response function of the detector and $Noise(x,y)$ stands for the noise from the background and the detector.\\

This paper primarily focuses on studying long-exposure images captured by ground-based optical telescopes, but our framework can be applied to images collected by any telescope, provided that we have an appropriate simulator. In the case of ground-based long exposure observations, atmospheric turbulence has a significant impact on the PSFs, which can be accurately modeled by using the Moffat model with $\beta$ equal to 4.765 \citep{1969A&A.....3..455M}. We set the Full-Width Half Maximum (FWHM) of the PSF as the first free parameter in the simulator. Additionally, we use the Gaussian random number to model various levels of noise caused by the detector and background, and we set the standard deviation as the second free parameter in the simulator. By using these parameters, we can generate images with realistic PSFs and noise, which better represent actual observation conditions for image restoration algorithms. Finally, we convolve several high-resolution template images with PSFs and add noise to generate the training data.\\

\subsection{The Image Restoration Part}
\label{subsec22}

\begin{figure}
	\centering
	\includegraphics[width=0.45\textwidth]{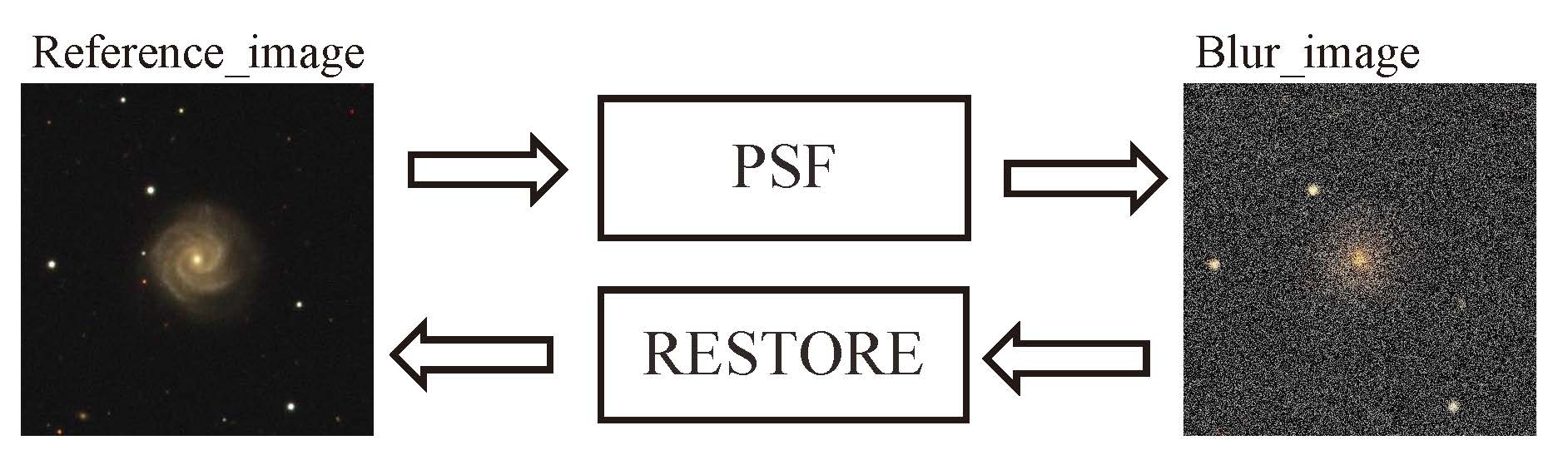}
	\caption{This figure shows the structure of the PSF-NET. It includes an image restoration neural network (RESTORE) and a PSF generation neural network (PSF). } 
	\label{fig2}%
	\end{figure}

\begin{figure*}
	\centering
	\includegraphics[width=0.98\textwidth]{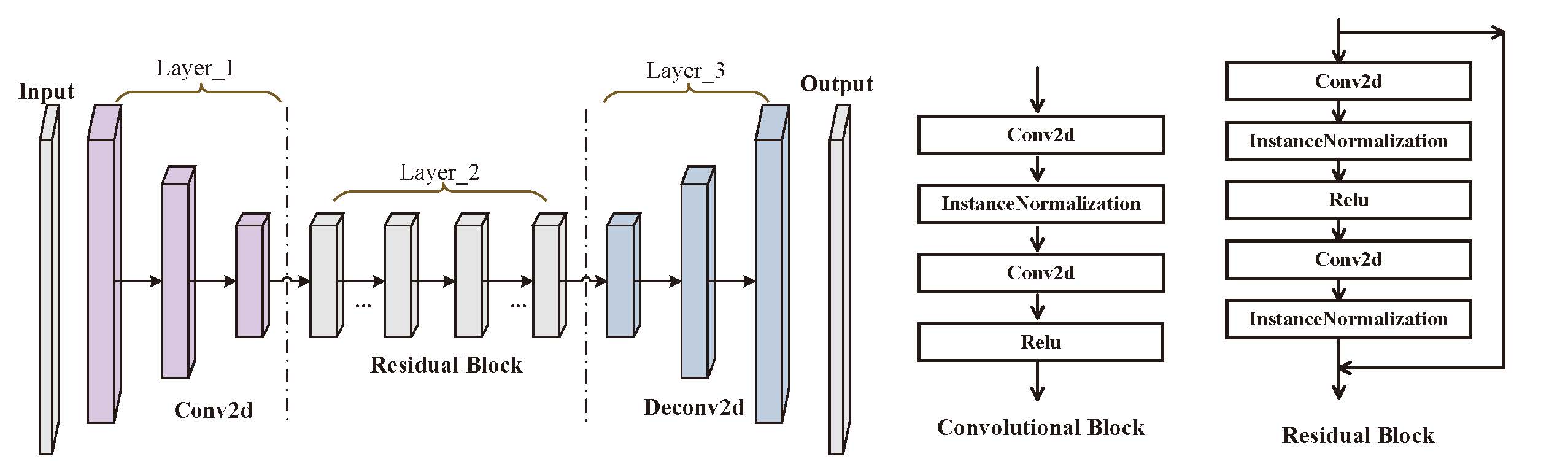}
	\caption{The left panel of the figure displays the architecture of the PSF neural network, which is similar to that of the RESTORE neural network. The middle panel illustrates the convolutional block, while the right panel depicts the residual block.}
	\label{fig3}%
	\end{figure*}

The image restoration neural network we use in this study is the PSF-NET, which was proposed by \citet{jia2020psf}. The PSF-NET consists of two neural networks, namely the PSF network (PSF) and the RESTORE network (RESTORE), as shown in the Figure~\ref{fig2}. Here is a detailed description of their roles and functionalities: Firstly, the primary task of the PSF network is to model the image's blurring process. It achieves this by learning and representing the characteristics of the PSF and noise, transforming a high-resolution image into a blurred version. The objective of this step is to capture the fundamental information regarding image blurring, thereby enhancing the accuracy and efficiency of the subsequent restoration process. Secondly, the RESTORE network plays the role of a deconvolution algorithm, tasked with generating high-resolution, clear images from the blurred images. After being trained, the RESTORE network is capable of effectively restoring blurred images to high-quality images, recovering the fine details and information in the image. It is noteworthy that both the PSF and RESTORE networks comprise 6 residual blocks, along with several convolutional and transposed convolutional blocks. The structures of these blocks are depicted in Figure \ref{fig3}. During the training phase, we adopt a joint training approach to train both the PSF network and the RESTORE network, thereby improving training efficiency and mitigating overfitting risks. After training, the RESTORE network can be deployed for the task of restoring blurred images.\\

An important point to consider is that both the PSF network and the RESTORE network are trained together, with the PSF network acting as a constraint for the RESTORE network. Unlike a simple deconvolution algorithm, the RESTORE network handles deconvolution and noise reduction. Therefore the PSF network learns impacts brought by the PSF and the noise. This structural design not only infuses the restoration process with enhanced physical realism but also bolsters the model's ability to generalize, enabling it to address various blurring scenarios effectively. The PSF neural network in this study generates blurred images from high-resolution images, while the Restore neural network generates high-resolution images from blurred images. To reflect the functions of these two neural networks, we have designed the loss function as presented in Equation~\ref{eq1}.
\begin{equation}
\label{eq1}
\begin{aligned}
	Loss = L_{idt}+L_{rec}+L_{fl},
	\end{aligned}
	\end{equation}
where $L_{idt}$ is the identity loss function, $L_{rec}$ is the cycle loss function and $L_{fl}$ is the focal frequency loss function. They are defined in equation~\ref{eq2},
\begin{eqnarray}
	\label{eq2}
	\begin{aligned}
		L_{idt} = \| PSF(Img_{org})-Img_{blur}\|_{2}+\\
			\|Restore(Img_{blur})-Img_{org}\|_{2}, \\
			L_{rec} = \|Restore(PSF(Img_{org}))-Img_{org}\|_{2}+\\
			\|PSF(Restore(Img_{blur}))-Img_{blur}\|_{2},\\
			L_{ffl} = W\times L_{fl},
	\end{aligned}
	\end{eqnarray}
where $PSF$ and $Restore$ stand for operators carried out by the PSF neural network and the RESTORE neural network, $Img_{blur}$ and $Img_{org}$ stand for blurred images generated by the simulation method, and original high resolution images. We use the $L_{idt}$ and the $L_{rec}$ to set the mean square error (MSE) between the original images and the images restored by the neural network as small as possible.\\
	
Both the PSF neural network and the RESTORE neural network adopt an encoder-decoder structure. However, since the decoder contains several up-sampling deconvolution layers that may introduce gaps in restored images, using only the aforementioned loss function may pose a problem. Our empirical observations suggest that gaps affect the structure of restored galaxy images and are more noticeable in the spatial frequency domain. To address this issue, we propose using the focal frequency loss $L_{ffl}$ to further enhance the performance of our neural network. The $L_{ffl}$ was originally introduced by \citet{jiang2020focal}, and involves the use of regularized weights $W$ to modulate the power spectral density, with $L_{fl}$ representing the mean squared error (MSE) between the original and restored images in the spatial frequency domain, as defined in equation~\ref{eq3},
\begin{eqnarray}
	\begin{aligned}
	\label{eq3}
	W = | FFT(Img_{org})-FFT(Restore(Img_{blur}))|^{\alpha},\\
	L_{fl} = \|FFT(Img_{org})-FFT(Restore(Img_{blur}))\|^2,
	\end{aligned}
	\end{eqnarray}
The fast Fourier transform is represented by $FFT$ and the regularized parameter $\alpha$ is used in the focal frequency loss $L_{ffl}$, which we defined as 1 in this study.\\

\subsection{The Parameter Selection Part}
\label{subsec23}
The parameter selection part is the central part of our framework, responsible for adjusting the proportion of images with various blur or noise levels used for training the neural network. Additionally, this part oversees the execution of the simulation and image restoration parts in parallel using the MPICH \citep{mpich2}. We will discuss the parameter selection component below.\\
	
To begin, our parameter selection part involves defining a dictionary to control the distribution of images with varying blur or noise levels in the training set. The keys of this dictionary are determined by the FWHM of PSFs and noise levels, while the values represent the mean loss function of all images with the respective FWHM or noise level in the test set. These values are updated at the end of each epoch. Subsequently, we generate a new set of images based on these updated key values using equation~\ref{eq4},
\begin{eqnarray}
	\begin{aligned}
	\label{eq4}
		test_{count}(r_i) = (total(r_i)-10) \times (1-list(r_i))+10,\\
		train_{count}(r_i) = (total(r_i)-10) \times list(r_i),
	\end{aligned}
	\end{eqnarray}
$test_{count} (r_i)$ and $train_{count} (r_i)$ represent the total number of images with a particular FWHM and noise level ($r_i$) in the test set and the training set, respectively. $total(r_i)$ represents the total number of images with the same FWHM or noise level. $list$ represents the parameter dictionary used to store the normalized mean value of the loss function for all images with predefined FWHMs of PSFs or noise levels. The values in the $list$ are used to calculate the percentage of images at a particular FWHM and noise level. By generating more blurred images that cannot be restored well by the restoration neural network in the previous epoch, the network is trained to gain better performance for these images in the next epoch because we have more images of that particular blur or noise level. With the parameter selection part, our neural network can effectively sample the space of blur levels or noise levels for a particular sky survey project and obtain a stable generalization ability for all images obtained by the project.\\

\begin{figure*}
	\centering
	\includegraphics[width=0.6\textwidth]{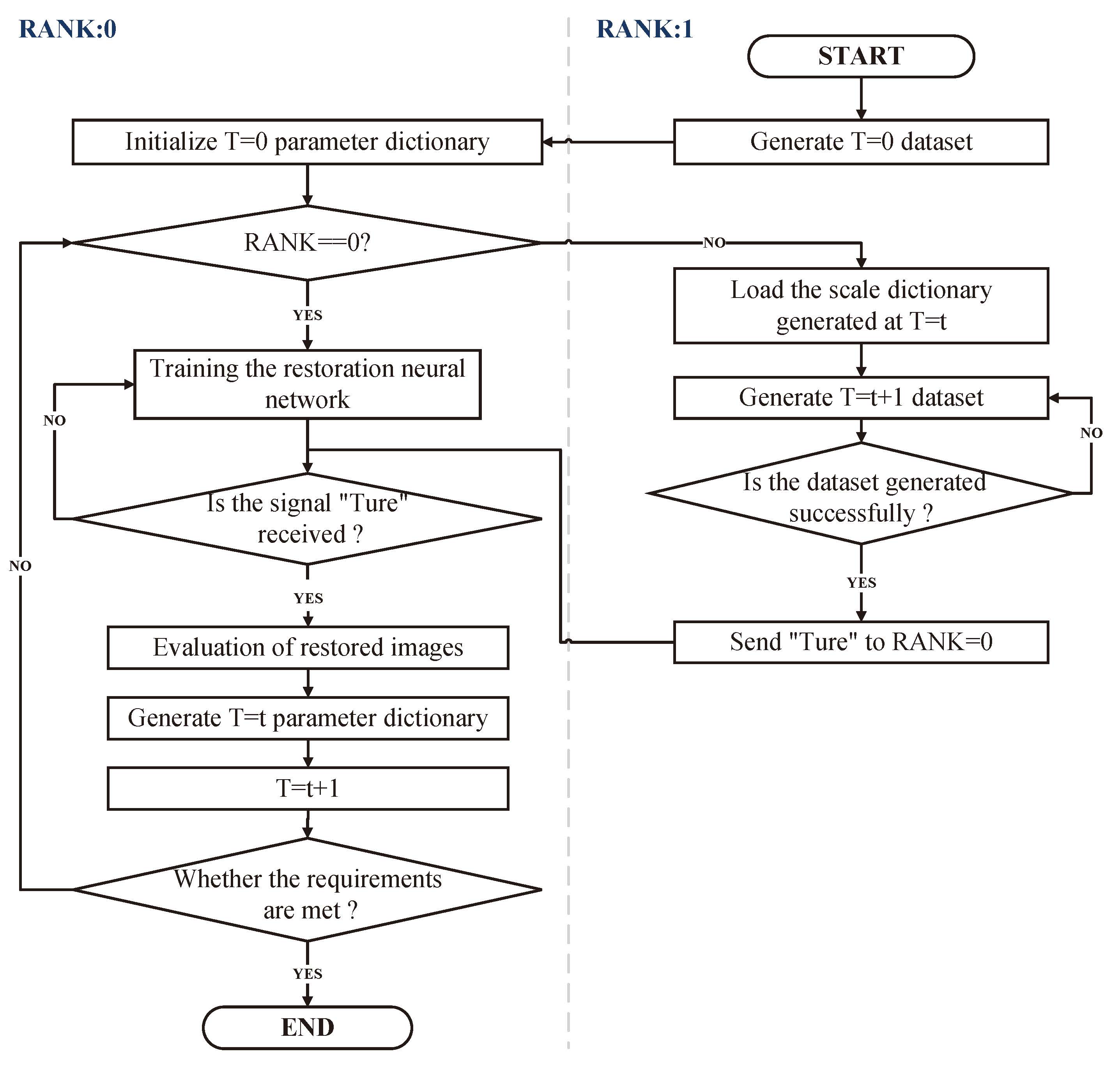}
	\caption{The flow chart of the parameter selection part is shown in this figure. The parameter selection part comprises two ranks, each with its own distinct role. Rank0 is responsible for training and testing the image restoration component, and runs primarily on a computer equipped with an RTX 3090 GPU. On the other hand, Rank1 is responsible for generating blurred images using the simulation component and runs on several other computers equipped with CPUs. For this purpose, we utilize older laptops or desktops that are equipped with i5 or i7 processors to run the simulation code.}
	\label{fig5}%
	\end{figure*}

To accelerate the computationally expensive simulation and image restoration parts, we propose dynamically adjusting the computation load on different computers during the training stage. The parameter selection part controls the data generation and image restoration parts and is illustrated in Figure~\ref{fig5}. The parameter selection part has two ranks. Rank0 trains and tests the image restoration part, which requires significant GPU resources on a single computer. Rank1 generates blur images with the data generation part on several other computers, which require CPU resources. The training data is set with an appropriate number based on the data generation and image restoration speeds. During the training stage, Rank1 generates blur images and sends a "TRUE" signal to Rank0 when all images in the training set of one epoch are generated. Rank1 continues to generate blurred images even after sending the "TRUE" signal. Meanwhile, Rank0 restores these blur images before receiving the "TRUE" signal from Rank1. Upon receiving the "TRUE" signal, Rank0 stops training and evaluates the qualities of all restored images. The parameter selection part calculates the MSE of all restored images in the test set and updates the parameter dictionary. The framework runs continuously with the parameter dictionary as input parameters until the set number of iterations is reached or the loss does not decrease for ten consecutive epochs.\\

\section{Performance Evaluation with Simulated and Real Observation Images}
\label{Sec:3}
To assess the effectiveness of our framework, we will test it on both simulated and real observational data. In Section~\ref{SubSec31}, we will introduce several criteria for evaluating its performance. In Section~\ref{SubSec32}, we will use simulated data to test the framework's performance. Since we have control over the blur and noise levels, we can effectively evaluate the neural network's generalization ability using simulated data. In Section~\ref{SubSec33}, we will evaluate the framework's performance using real observational data obtained from the SDSS project \citep{SDSSDR7}. We send images of galaxies directly to the neural network and the results demonstrate the effectiveness of our framework. Further details about our framework will be discussed below.\\

\subsection{Performance Evaluation Criteria}
\label{SubSec31}
We employ the peak signal-to-noise ratio (PSNR) as defined by \citet{xu2014deep} and the structural similarity (SSIM) as defined by \citet{wang2004image} to quantitatively assess the quality of the images. The equation for PSNR can be found in equation~\ref{psnr}:
\begin{eqnarray}
	\begin{aligned}
	\label{psnr}
	PSNR (img,img_{ref}) = 20\log_{10} \left(\frac{MAX(img_{ref})}{\sqrt{MSE}}\right), \\
        MSE = \frac{1}{\sqrt{N \times M}}\sum_{i=1}^{N} \sum_{j=1}^{M}\left(img_{(i,j)}-img_{ref(i,j)}\right)
	\end{aligned}
\end{eqnarray}
where $N\times M$ is the size of the image, $img$ and $img_{ref}$ are restored images and original images. For simulated images, we could obtain the $img$ and $img_{ref}$ and for real observation images, $img_{ref}$ is mean values of $img$. The PSNR can directly reflect the similarity between restored images and blurred images. Images with larger PSNR would have better quality. Meanwhile, the SSIM is defined in the equation~\ref{ssim}:
\begin{eqnarray}
\label{ssim}
\begin{aligned}
   SSIM(img ,img_{ref}) & = l(img,img_{ref})  \\
  & \cdot c(img, img_{ref}) \cdot s(img,img_{ref}), \\
  l(img,img_{ref}) & =  \frac{2\mu_{img}\mu_{ref}+C_1}{\mu^{2}_{img}+\mu^{2}_{ref}+C_1}, \\
   c(img,img_{ref}) & =  \frac{2\delta_{img}\delta_{ref}+C_2}{\delta^{2}_{img}+\delta^{2}_{ref}+C_2},\\
   s(img,img_{ref}) & =  \frac{\delta_{img,ref}+C_3}{\delta_{img}\delta_{ref}+C_3},\\
  C_1 & = (K_1L)^2 ,\\
  C_2 & = (K_2L)^2, \\
  C_3 & = C_2/2,
    \end{aligned}
\end{eqnarray}
 where $\mu$ and $\delta$ are the average and standard deviation of the gray scale values, respectively, $\delta_{img, ref}$ is the covariance of the gray scale values, L is the dynamic range of the image, and $K_1$ and $K_2$ are small arbitrary values (0.001 in this paper). The SSIM is a perceptual model that considers the brightness, contrast, and intensity scale of two images simultaneously. The quality of an image is considered better if its SSIM is larger.\\

\subsection{Performance Evaluation With Simulated Images}
\label{SubSec32}
In order to evaluate the effectiveness of our framework on simulated images, we first simulate an observation scenario carried out by a ground-based telescope with long exposure. We obtain galaxy images in the r band from the SDSS DR 7 and generate blurred images through simulation. Our simulator assumes that the PSF follows the Moffat model, with a full-width at half-maximum (FWHM) distribution ranging from 2.0 to 8.0 pixels (equivalent to 0.792 arcsec to 3.168 arcsec) across 5 different levels, and Gaussian noise distributed equally across 5 different levels with $\sigma$ ranging from 1.0 to 15.0. This results in a total of 25 different levels of blurred images. We train our framework using the images generated by the simulator with the aforementioned parameters. Additionally, we generate blurred images with higher levels of blur and noise to evaluate the performance of our algorithm. We divide the FWHM of the PSFs into 13 bins ranging from 2 to 14 pixels (equivalent to 0.792 arcsec to 5.544 arcsec), with the FWHMs of the PSFs in each bin considered as random variables. We also divide the noise levels into 24 bins ranging from 2 to 25, with $\sigma$ considered as a random variable in each bin.\\

\begin{figure*}
	\centering
	\includegraphics[width=0.40\textwidth]{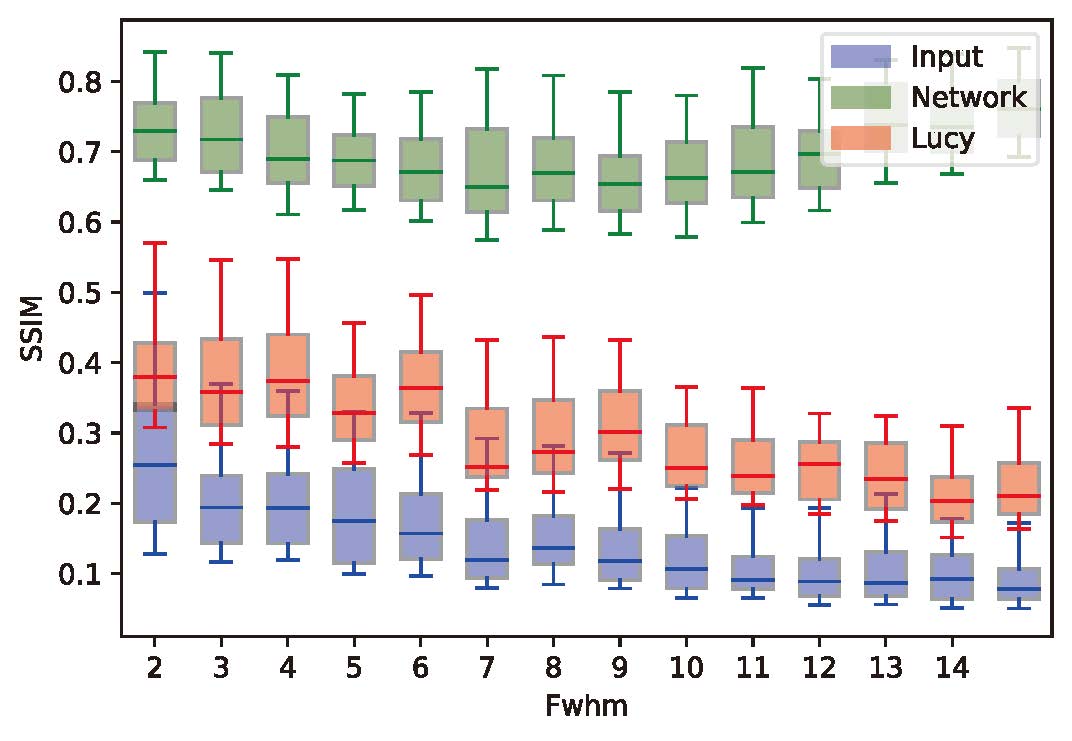}
         \includegraphics[width=0.40\textwidth]{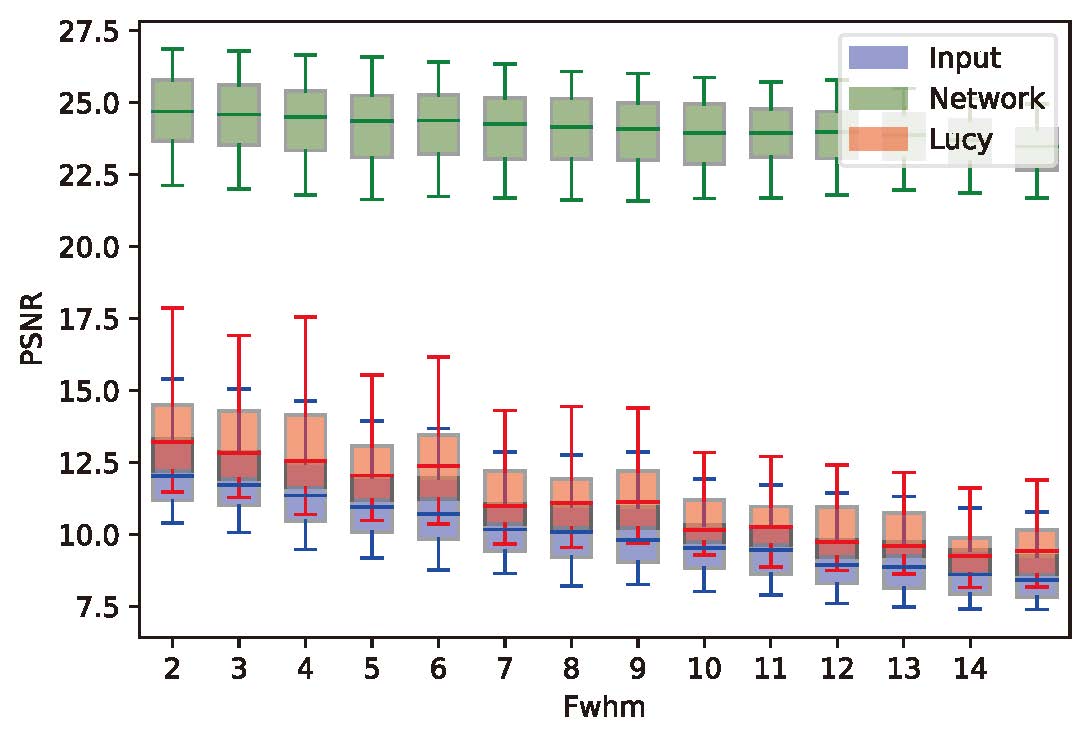}\\
         \includegraphics[width=0.40\textwidth]{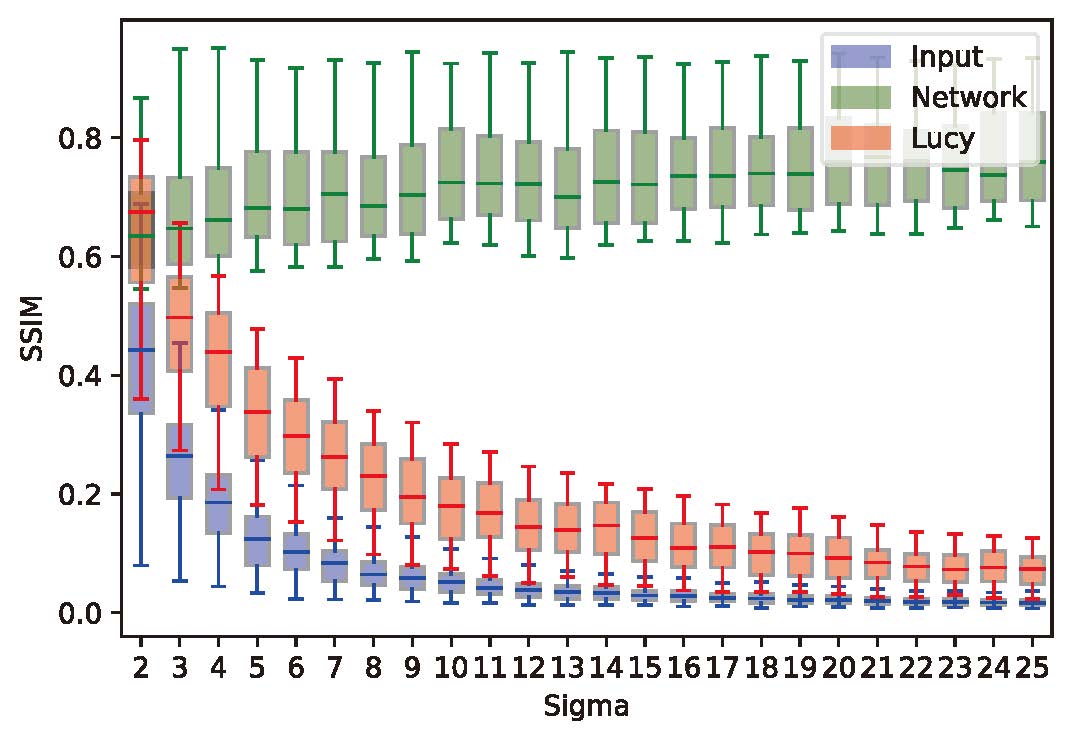}
         \includegraphics[width=0.40\textwidth]{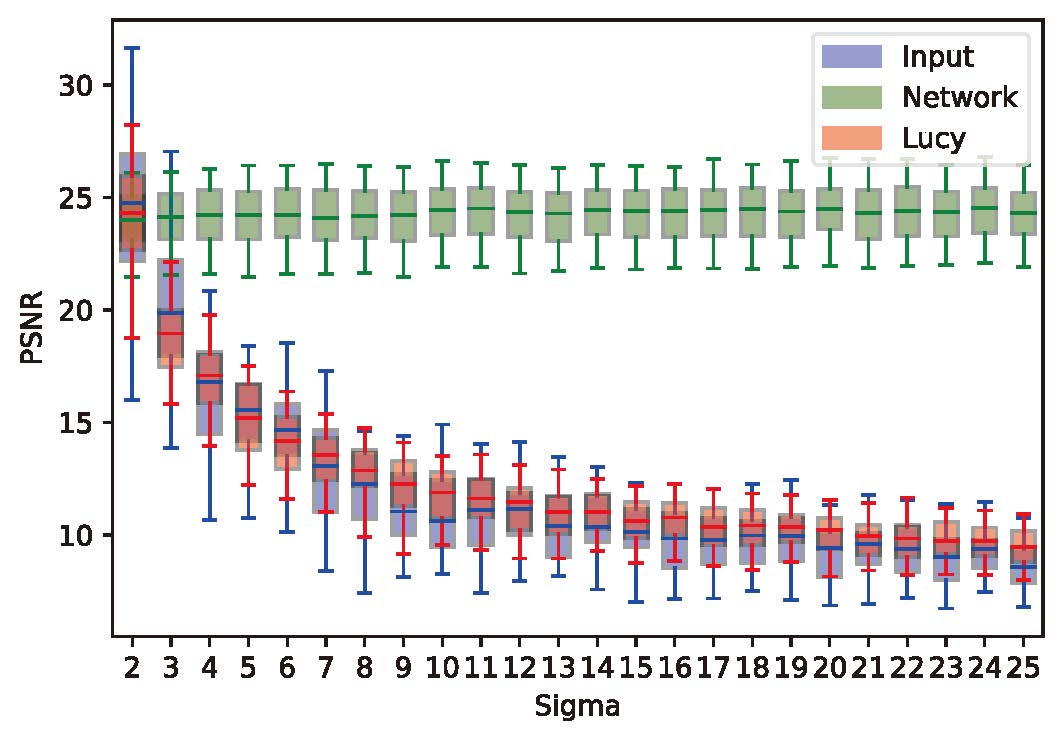}\\
	\caption{Comparison of different mage restoration algorithms under different Noise and PSF Conditions. The box plot showing the differences between our restoration algorithm (Green) and the RL (Red) method under different PSF conditions (Top Panel) and different noise levels (Bottom Panel). }
	\label{box}%
\end{figure*}

In this section, we apply the evaluation criteria outlined in Section~\ref{SubSec31} to assess the effectiveness of our framework. We utilize box plots to illustrate the performance of our algorithm and the Richardson-Lucy method (RL) with the Gaussian denoising method across varying noise levels and different PSFs. In Figure~\ref{box}, results obtained from our algorithm are denoted in green, RL algorithm results in red, and the original blurred data in blue. Upon analyzing these box plots, it becomes apparent that our algorithm consistently outperforms the RL method under diverse conditions, encompassing various noise levels and distinct PSFs. Notably, our algorithm exhibits a significant advantage when dealing with datasets characterized by fluctuating noise levels, demonstrating its robustness and stability in the presence of noisy input data.\\

Furthermore, we conduct a t-test to determine whether a significant performance difference exists between the RL algorithm and our model. Under the null hypothesis (H0), we assume there is no performance difference between the RL algorithm and our model. The alternative hypothesis (H1) posits a performance difference between the RL algorithm and our model. In this study, we conduct statistical significance analyses for different scenarios and calculate p-values, all of which are less than 0.05, as shown in Tables~\ref{table1}, \ref{table2}, \ref{table3}, and \ref{table4}. In all these cases, we reject the null hypothesis, signifying a significant performance difference between the RL algorithm and our model. These findings collectively underscore the remarkable image restoration quality achieved by our algorithm. When evaluated using PSNR or SSIM metrics, our algorithm consistently outperforms the traditional RL method. In summary, our test results unequivocally establish the superior performance of our image restoration framework across various observation conditions.\\


\begin{table*}
  \centering
    \begin{tabular}{|l|c|c|c|c|c|c|c|}
        \hline
        \textbf{Fwhm} & 2 & 4 & 6 & 8 & 10 & 12 & 14 \\
        \hline
        \textbf{t$\_$statistic} & 17.97 & 20.46 & 17.57 & 20.27 & 20.76 & 18.79 & 24.56 \\
        \hline
        \textbf{p$\_$value} & 5.96e-38 & 1.23e-43 & 5.08e-37 & 3.40e-43 & 2.72e-44 & 7.41e-40 & 3.21e-52 \\
        \hline
    \end{tabular}
    \caption{Statistical Analysis of SSIM Performance for RL and Our Algorithm under Different Blur Levels (PSFs).}
    \label{table1}
\end{table*}

\begin{table*}
  \centering
    \begin{tabular}{|l|c|c|c|c|c|c|c|c|}
        \hline
        \textbf{Fwhm} & 2 & 4 & 6 & 8 & 10 & 12 & 14 \\
        \hline
        \textbf{t$\_$statistic} & 49.69 & 73.94 & 56.31 & 75.21 & 76.49 & 68.05 & 110.96 \\
        \hline
        \textbf{p$\_$value} & 6.02e-90 & 6.82e-113 & 4.29e-97 & 6.88e-114 & 7.09e-115 & 4.77e-108 & 8.31e-137 \\
        \hline
    \end{tabular}
    \caption{Statistical Analysis of PSNR Performance for RL and Our Algorithm under Different Blur Levels (PSFs).}
    \label{table2}
\end{table*}

\begin{table*}
  \centering
    \begin{tabular}{|l|c|c|c|c|c|c|c|c|}
        \hline
        \textbf{Sigma} & 2 & 9 & 12 & 15 & 18 & 21 & 23 & 25 \\
        \hline
        \textbf{t$\_$statistic} & 1.70 & 17.81 & 26.46 & 32.40 & 32.12 & 40.43 & 40.60 & 43.98 \\
        \hline
        \textbf{p$\_$value} & 4.88e-02 & 3.25e-29 & 1.05e-40 & 5.35e-47 & 1.00e-46 & 4.40e-54 & 3.24e-54 & 1.63e-54 \\
        \hline
    \end{tabular}
  \caption{Statistical Analysis of SSIM Performance for RL and Our Algorithm under Different Noise Levels.}
  \label{table3}
\end{table*}

\begin{table*}
  \centering

    \begin{tabular}{|l|c|c|c|c|c|c|c|c|}
        \hline
        \textbf{Sigma} & 2 & 9 & 12 & 15 & 18 & 21 & 23 & 25 \\
        \hline
        \textbf{t$\_$statistic} & 6.66 & 34.76 & 42.52 & 39.62 & 43.38 & 50.09 & 47.96 & 59.78 \\
        \hline
        \textbf{p$\_$value} & 3.52e-09 & 3.11e-49 & 1.02e-55 & 1.98e-53 & 2.27e-56 & 4.52e-61 & 1.19e-59 & 6.52e-67 \\
        \hline
    \end{tabular}
      \caption{Statistical Analysis of PSNR Performance for RL and Our Algorithm under Different Noise Levels.}
      \label{table4}
\end{table*}

\begin{figure*}
	\centering
	\includegraphics[width=0.95\textwidth]{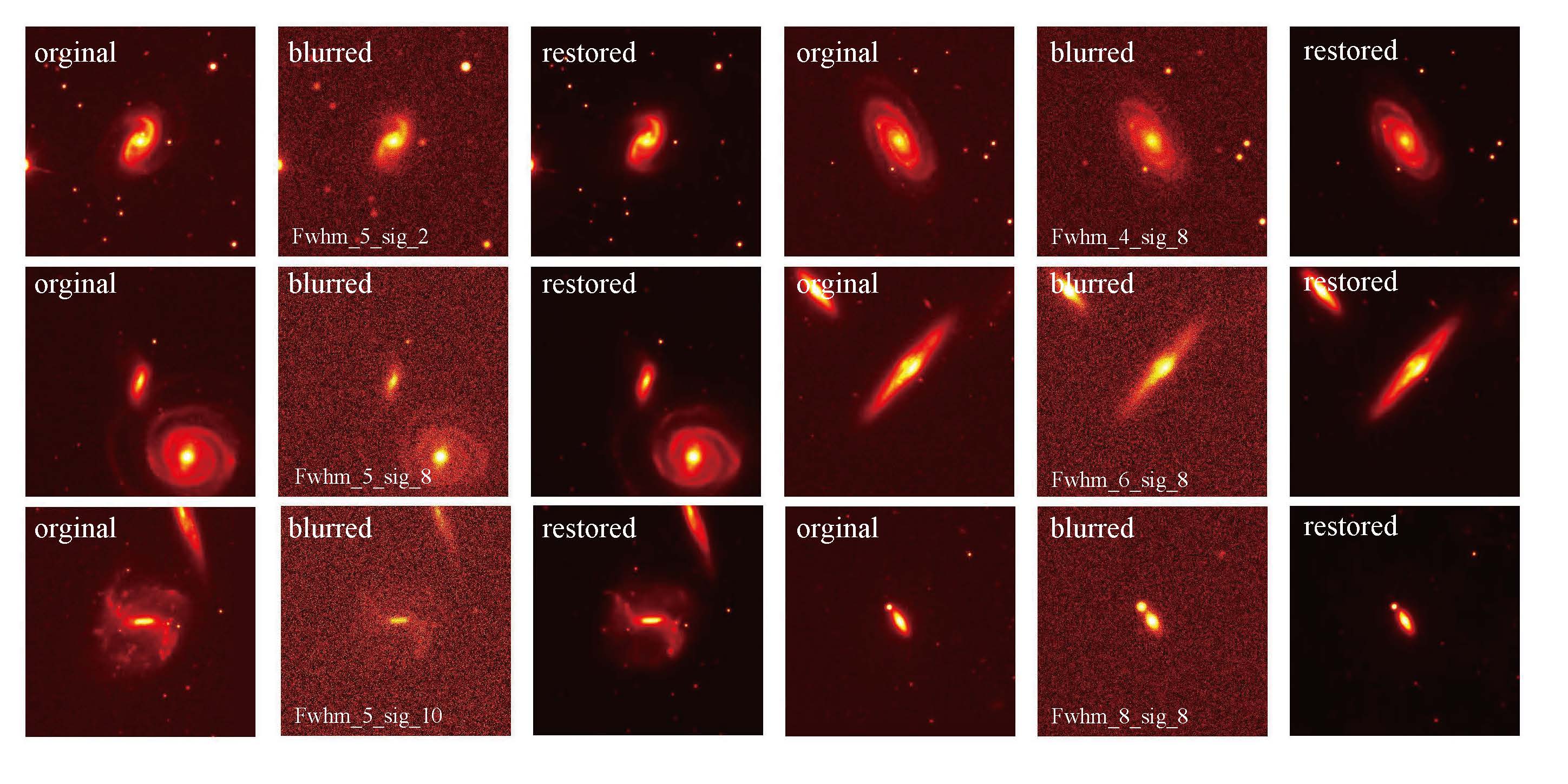}
	\caption{The performance of our framework in the restoration of images that are generated within FWHMs or noise levels defined by the training set.}
	\label{qualitative-1}%
\end{figure*}

\begin{figure*}
	\centering
	\includegraphics[width=0.93\textwidth]{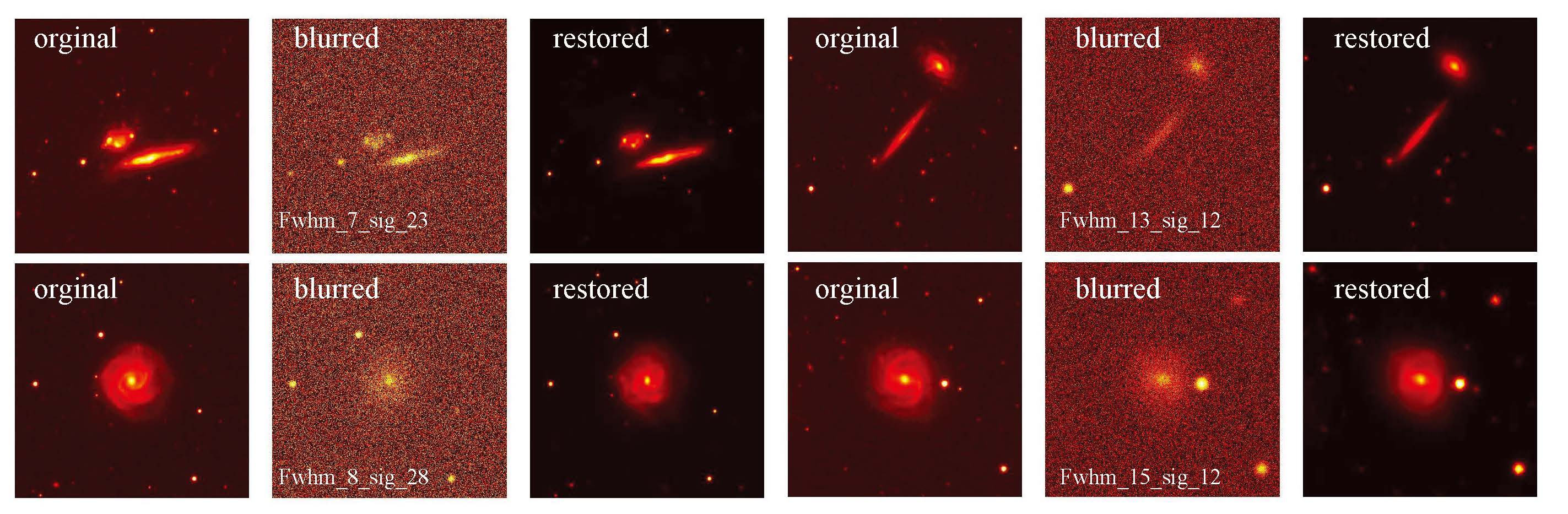}
	\caption{The performance of our framework in the restoration of images that are generated with larger FWHM or higher noise level defined by the training set.}
	\label{qualitative}
\end{figure*}

To assess the performance of our algorithm in a qualitative manner, we utilize simulated images to test the effectiveness of our framework. Specifically, we apply the trained RESTORE neural network to restore simulated images that have PSFs and noise levels within the range defined by the training set. The resulting images are presented in Figure~\ref{qualitative-1}. The figure illustrates that the trained RESTORE neural network can significantly enhance the quality of the images. The fine details, such as spirals and bars of galaxies, can be clearly observed in the restored images. Additionally, we test the performance of our framework on simulated images that have PSFs or noise levels that exceed the range defined in the training set. The results are shown in Figure~\ref{qualitative}, and it is evident that the trained RESTORE neural network exhibits a strong generalization ability, even for images with larger FWHM or higher noise levels.\\

\subsection{Performance Evaluation With Real Observation Images}
\label{SubSec33}
Frist, we demonstrate the efficacy of our framework with real observation data obtained from the SDSS Data Release 7 \citep{sdss2000, SDSSDR7}. We employ our method to restore images of low surface brightness galaxies (LSBGs) as detected by \citet{yi2022automatic}. LSBGs are a class of galaxies with central surface brightness fainter than the sky background, often exhibiting high gas content and believed to be in the early stages of galaxy formation or to have recently undergone a burst of star formation \citep{impey1997low}. Due to their low luminosity, studying LSBGs is challenging \citep{du2015low}, making image restoration algorithms necessary to enhance their image quality before any scientific analysis can be performed. However, it is difficult to use traditional deconvolution-based image restoration algorithms, which require bright stars as references, and content-based image restoration methods are ineffective due to very small number of photons in LSBGs. Thus, we use our framework to restore LSBG images.\\

\begin{figure*}
	\centering
	\includegraphics[width=0.80\textwidth]{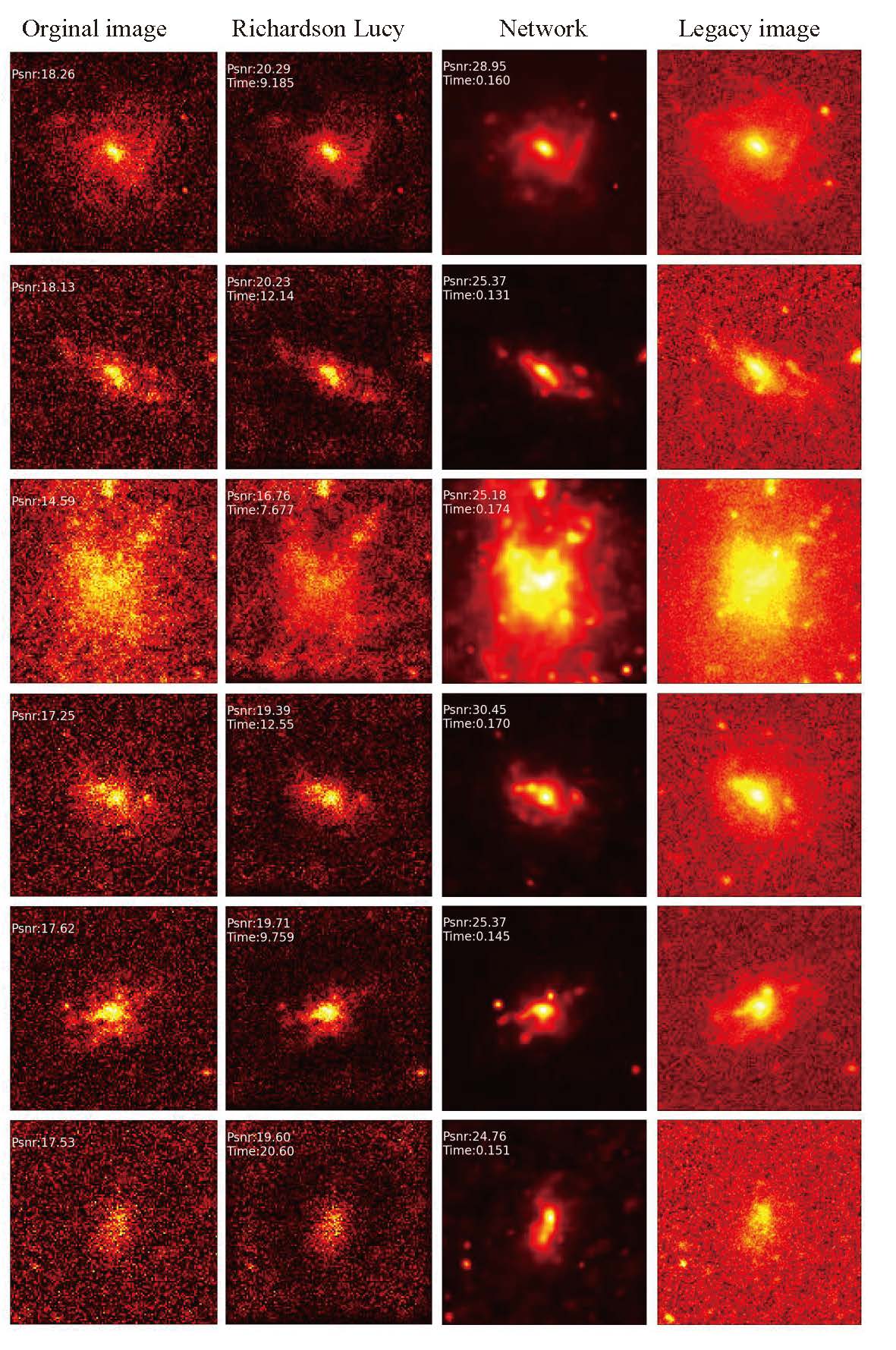}
	\caption{This figure shows original images obtained by the SDSS project in the r band, images restored by the RL method, with references of PSFs provided by \citet{Sainz2020mnras}, images restored by our framework and images of the same LSBG obtained by the DESI Legacy Imaging Surveys. In the upper left corner of each figure, we show the PSNR of restored images and calculation speed of different methods in restoration of different images. As shown in this figure, our framework could effectively restore blurred images.}
	\label{low}%
\end{figure*}

These images in the SDSS project are captured by a 2.5-meter telescope and have a pixel scale of 0.396 arcsec with an exposure time of 53.9 sec. As a result, the PSF and noise levels used in the simulation model in Section~\ref{subsec22} can reflect the properties of real observation data. Therefore, we train our framework with the simulator and the aforementioned parameters to obtain the weights of the RESTORE neural network. Subsequently, we apply the RESTORE neural network directly to process the LSBG images captured by the SDSS project in the r band. Figure~\ref{low} exhibits both the original images and the images restored by our neural network. Additionally, we also employ the RL deconvolution algorithm \citep{fish1995blind} and select PSF references according to \citet{Sainz2020mnras} to restore these images for comparison. Furthermore, we present the LSGB images acquired by the DESI Legacy Imaging Surveys \citep{dey2019overview} in this figure to provide a reliable reference. Since larger telescopes are used to execute the DESI Legacy Imaging Surveys, the data obtained from it can improve our evaluation of the effectiveness of our method.\\

Firstly, as demonstrated in the upper left corner of each figure, our framework can efficiently improve the PSNR of LSBGs with less time (around 10 times faster than the RL method with appropriate prior PSFs). Next, we examine the restored images in detail. Although the RL method could provide effective results, thanks to the accurate PSF model, images restored by the RL method are still affected by strong noises. Our framework effectively restores fine structures of these galaxies, such as spirals, disks, and filaments, and also reduce effects brought by noises. When comparing the restored images with those obtained by the DESI Legacy Imaging Surveys, we find that structures restored by our framework are true. It is worth noting that our framework builds PSFs and the deconvolution procedures for image restoration, so in principle it does not generate artificial features. Moreover, our framework can effectively suppress noise in these images, resulting in some images with even better quality than those obtained by the DESI Legacy Imaging Surveys. Overall, our framework can assist scientists in studying the properties and morphological structures of LSBGs in greater detail.\\

\begin{figure*}
	\centering
	\includegraphics[width=0.48\textwidth]{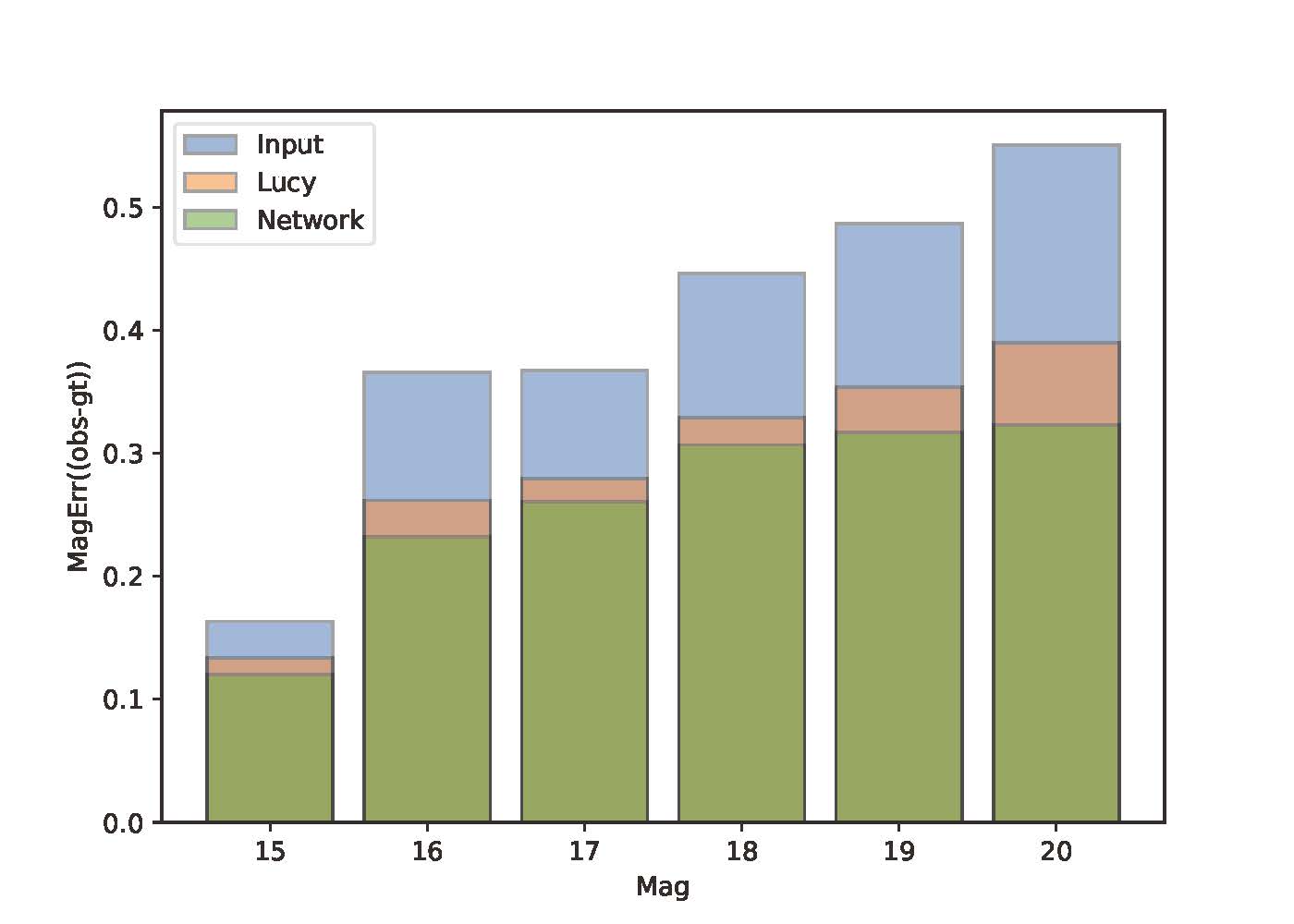}\\
        \includegraphics[width=0.3\textwidth]{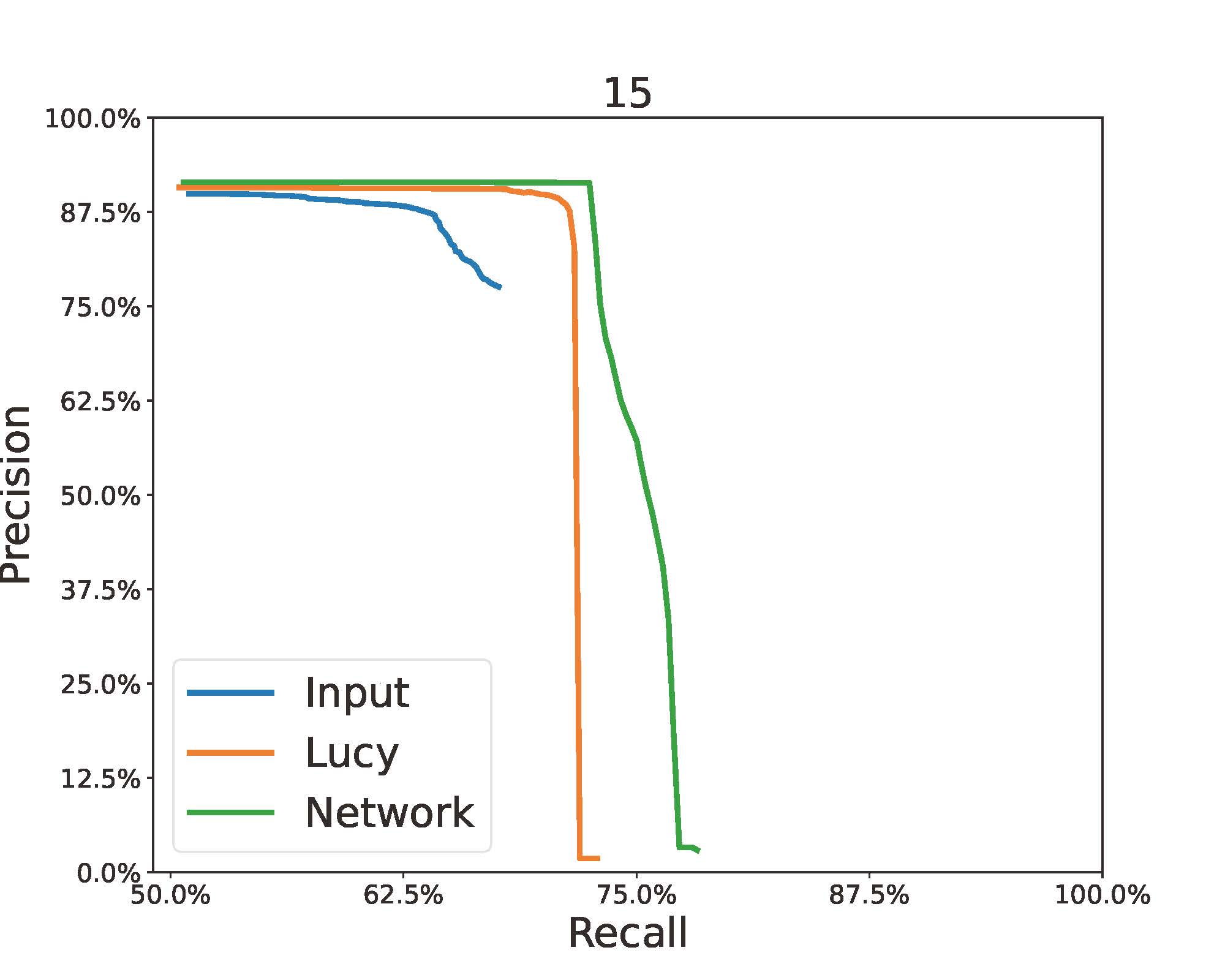}
        \includegraphics[width=0.3\textwidth]{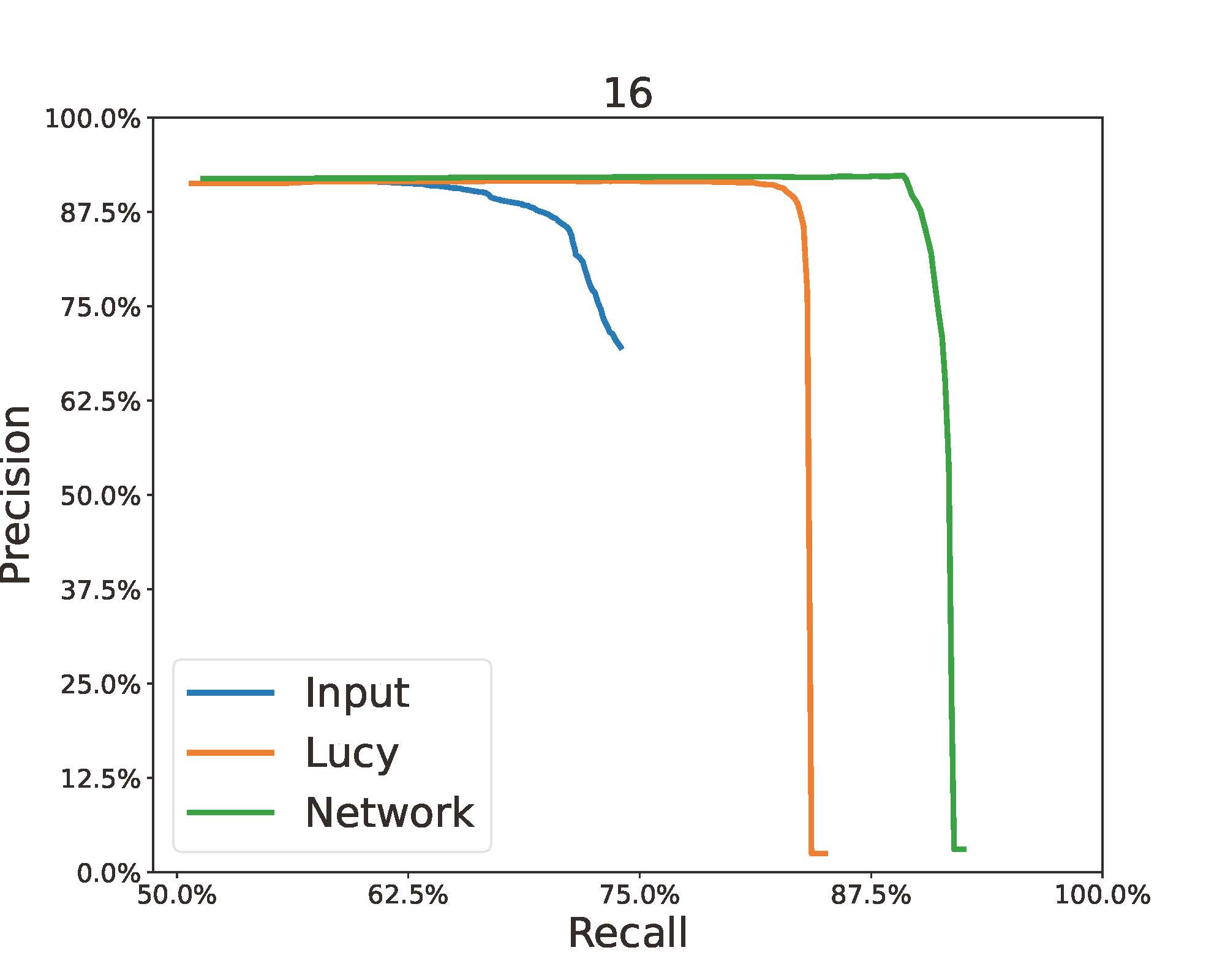}
        \includegraphics[width=0.3\textwidth]{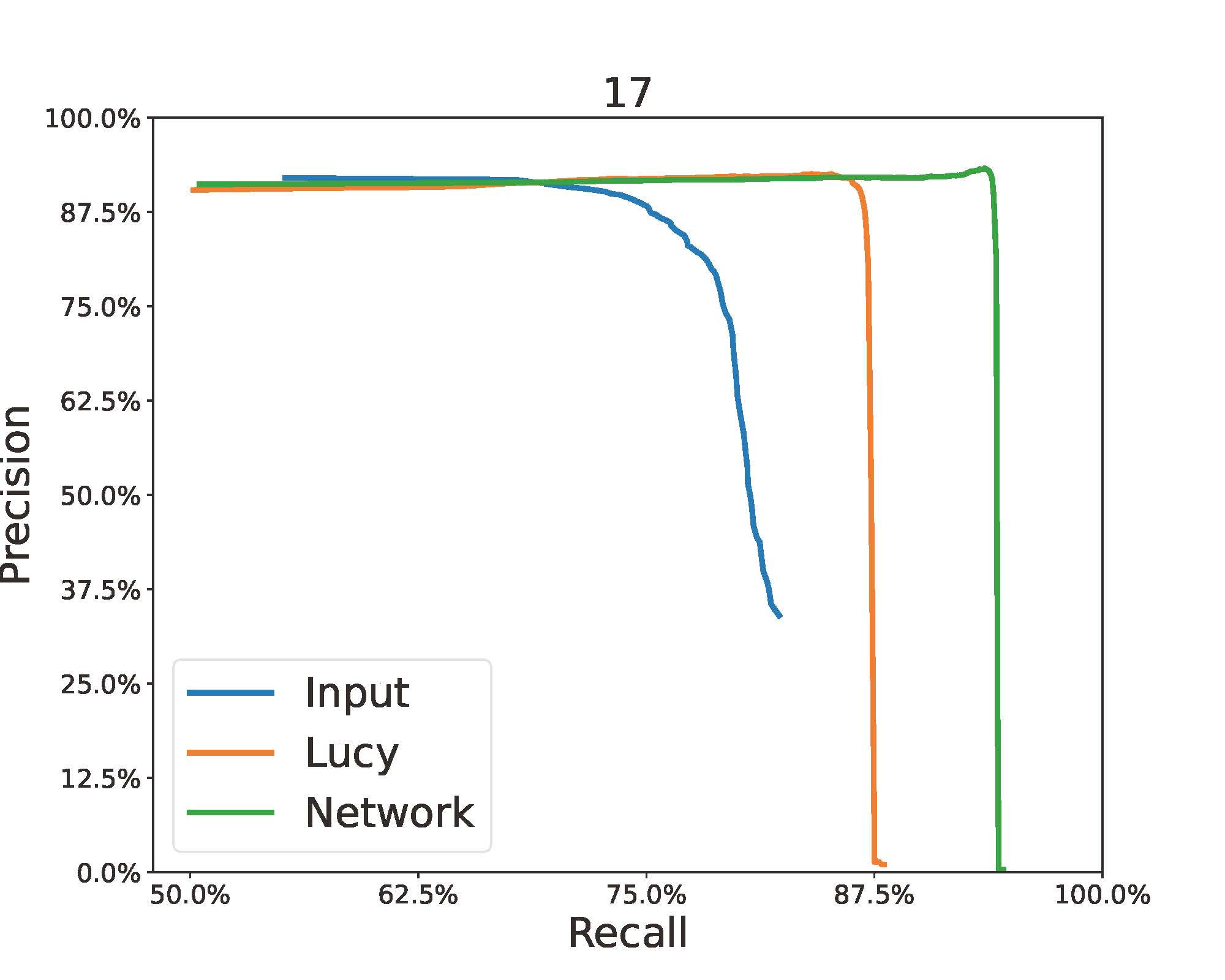}\\
        \includegraphics[width=0.3\textwidth]{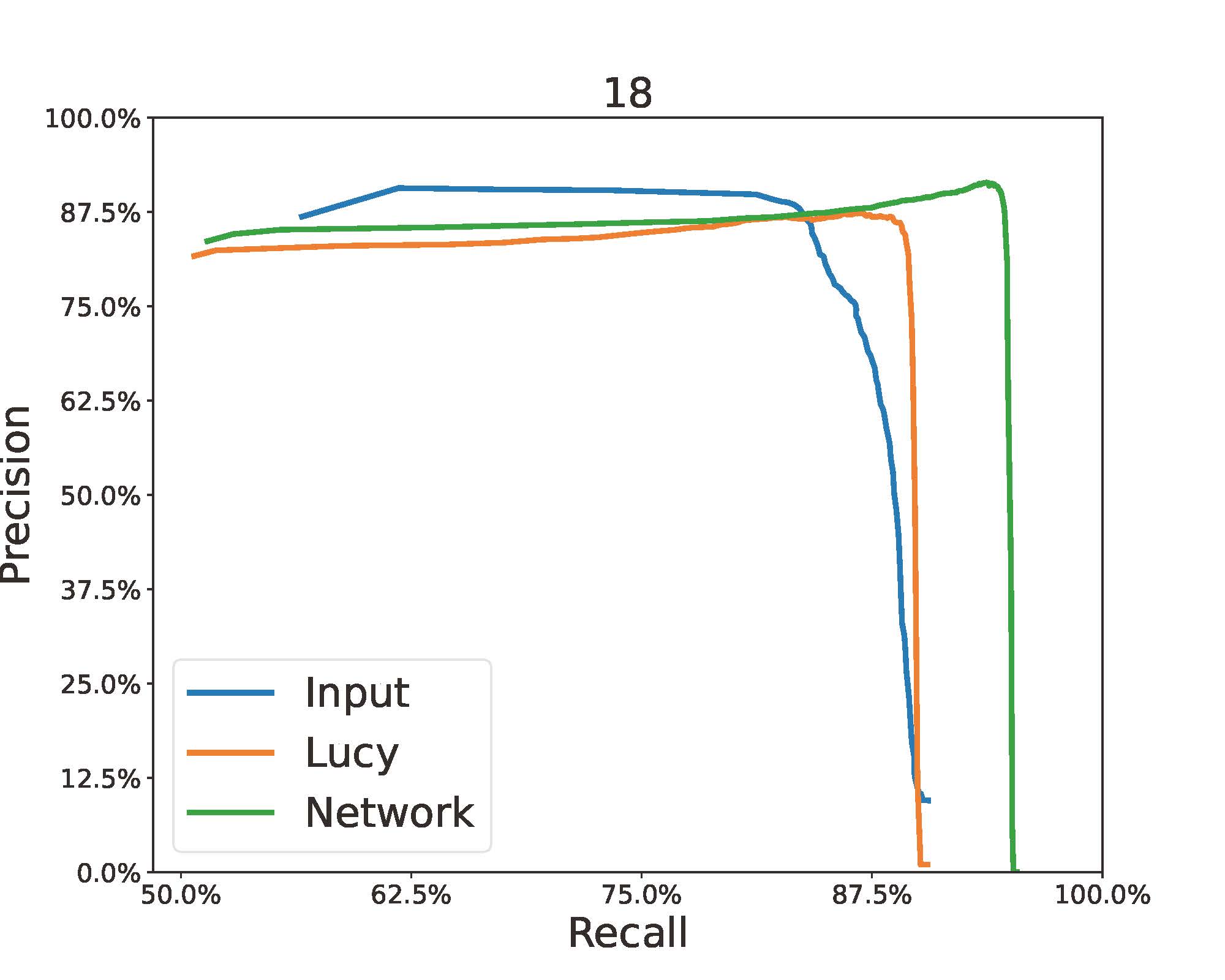}
        \includegraphics[width=0.3\textwidth]{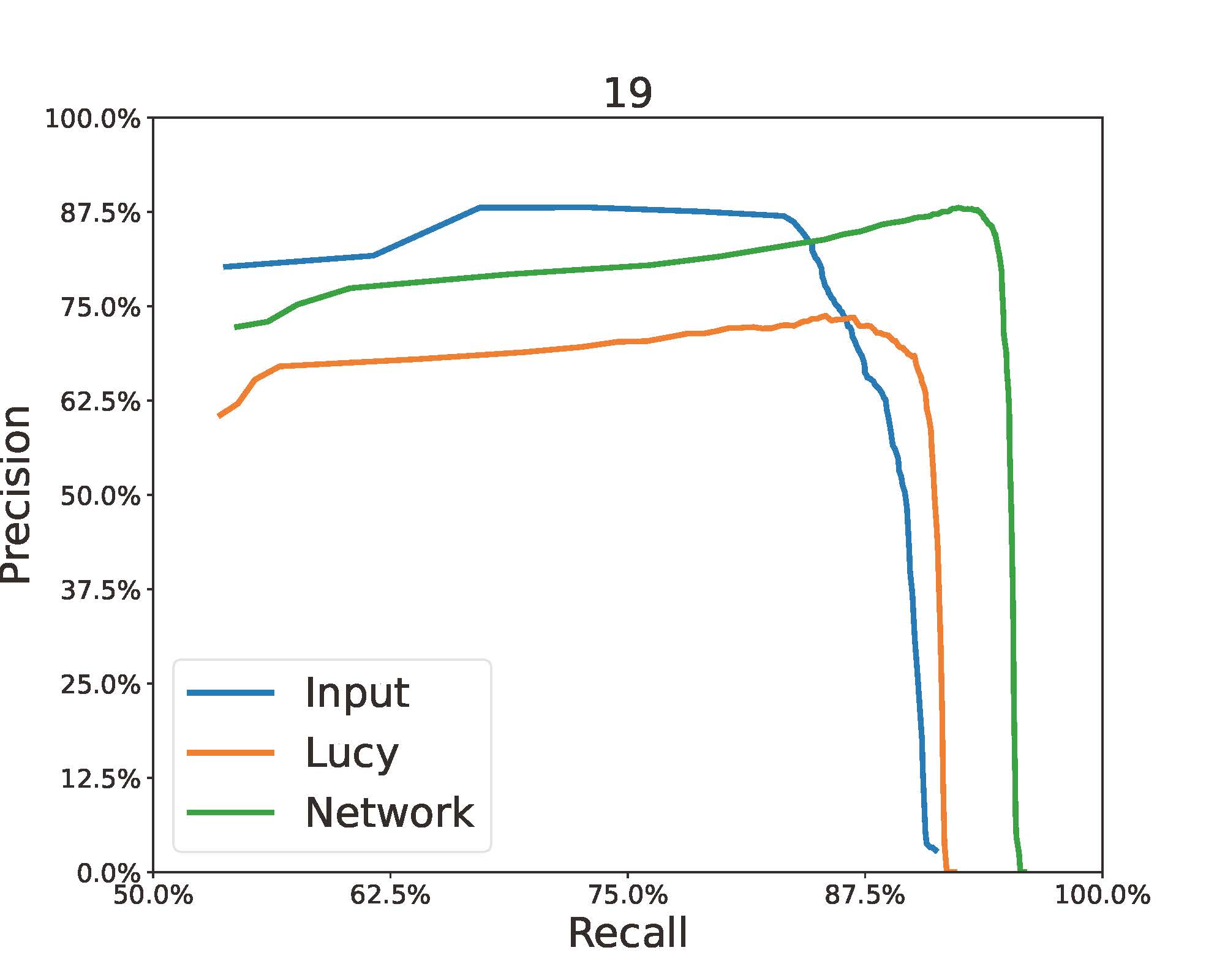}
        \includegraphics[width=0.3\textwidth]{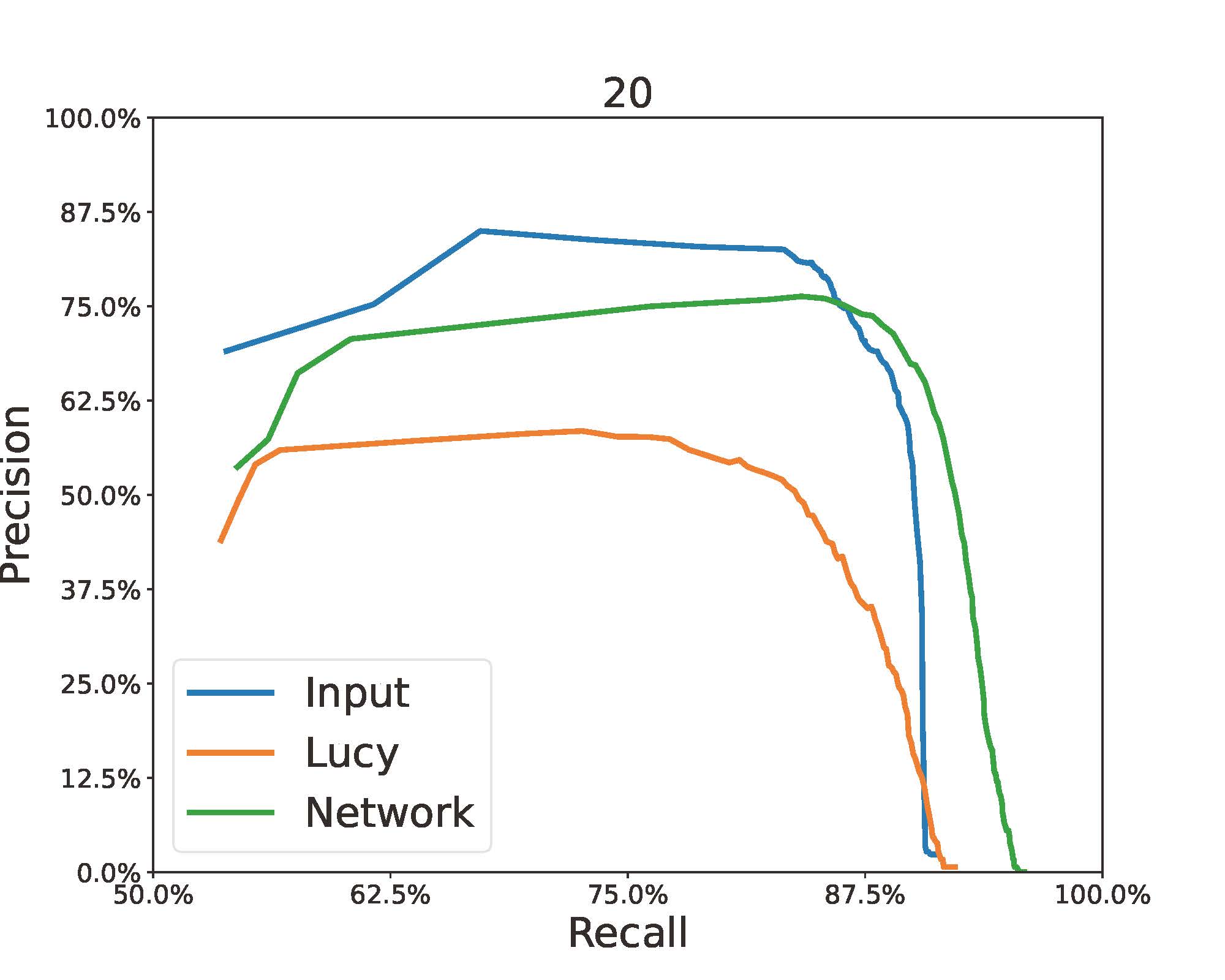}
	\caption{Comparison of photometry and detection results for stars with different magnitudes. MagErr: The average percentage error between photometric measurements and true values. P-R Curve: The Precision-Recall (P-R) curve is a graphical representation of the trade-off between precision and recall for a binary classification model. Precision is the ratio of true positive predictions to the total number of positive predictions, while recall is the ratio of true positive predictions to the total number of actual positive instances.}
	\label{qualitative-2}
\end{figure*}

In our further investigation, our primary focus was on evaluating the algorithm's performance concerning the enhancement of photometry accuracy and detection efficiency. To do this, we randomly select SDSS R band images, each with dimensions of 1024x1024 pixels. Subsequently, we apply both our image restoration algorithm and the RL algorithm to restore these images. Following restoration, we utilize the SExtractor for detection and photometry \citep{bertin1996sextractor}. The results of this study are visually presented in Figure~\ref{qualitative-2}. We conduct this analysis for each magnitude, considering a dataset of 2000 images for evaluation. Subsequently, we assess these results through a rigorous statistical analysis. Our findings indicate that the processed data, particularly the data treated with our algorithm, show a higher recall rate and precision rate. This outcome demonstrates that our algorithm significantly improves the efficiency of celestial object detection, especially for stars with low signal-to-noise ratio. Moreover, our method has proven effective in enhancing photometry accuracy, as vividly depicted in the top panel of Figure~\ref{qualitative-2}. These results provide strong support for the practical applicability of our algorithm in the field of astronomy.\\


\section{Conclusions and future works}
\label{Sec:4}
We have introduced a novel framework in this paper for restoring blurred astronomical images by merging deep learning techniques with simulation algorithms. Our framework actively trains the restoration neural network using a simulation algorithm that represents a specific telescope. Once trained, the restoration neural network can produce restored images more efficiently compared to the traditional RL deconvolution algorithm. We have tested our method on both simulated and real observational data and have found that it effectively minimizes the impact of noise and PSFs, making previously unseen fine structures of galaxies visible.\\

We have also identified two areas for future improvement. Firstly, we have shown that the physical parameters and prior model used to represent the image degradation process are crucial for our framework. To extend our approach to data from different sky survey projects, we need to develop an adequate parametric point spread function (PSF) model and an adequate telescope simulator. Therefore, we will introduce physics-informed machine learning algorithms as necessary tools to build PSF models. Meanwhile, the digital twin technology is a promising method for generating simulation data according to telemetry data and high-fidelity simulators, and we are currently developing a digital twin as the telescope simulator \citep{jia2022digital,zhan2022database}. Secondly, we use the $L_2$ norm and FFL to train our restoration neural network, and we could further investigate new regularization methods based on human attention and big data obtained from previous sky survey projects to improve the performance of our framework.\\

Overall, our proposed framework is suitable for restoring images obtained from future sky survey projects, such as the LSST, the Euclid, and the CSST. Our framework could help scientists to better recognize the morphology of galaxies, which could increase outcomes from citizen science platforms. Besides, our framework could better increase accuracy of shape measurements for galaxies with low signal to noise ratio. We are now using our framework to process data obtained by the DECaLS \citep{DESC} for further scientific research. We will deploy our method for data obtained by the CSST and the Euclid in the future. \\

\section*{Acknowledgements}
This work is supported by National Natural Science Foundation of China (NSFC) with funding number of 12173027 and 12173062. We acknowledge the science research grants from the China Manned Space Project with NO. CMS-CSST-2021-A01. We acknowledge the science research grants from the Square Kilometer Array (SKA) Project with NO. 2020SKA0110102. Peng Jia acknowledges support from the Civil Aerospace Technology Research Project (D050105) and the Major Key Project of PCL. Jiameng Lv acknowledges support from the Shanxi Graduate Innovation Project (2022Y274).\\

Funding for the SDSS and SDSS-II has been provided by the Alfred P. Sloan Foundation, the Participating Institutions, the National Science Foundation, the U.S. Department of Energy, the National Aeronautics and Space Administration, the Japanese Monbukagakusho, the Max Planck Society, and the Higher Education Funding Council for England. The SDSS Web Site is http://www.sdss.org/.\\

The SDSS is managed by the Astrophysical Research Consortium for the Participating Institutions. The Participating Institutions are the American Museum of Natural History, Astrophysical Institute Potsdam, University of Basel, University of Cambridge, Case Western Reserve University, University of Chicago, Drexel University, Fermilab, the Institute for Advanced Study, the Japan Participation Group, Johns Hopkins University, the Joint Institute for Nuclear Astrophysics, the Kavli Institute for Particle Astrophysics and Cosmology, the Korean Scientist Group, the Chinese Academy of Sciences (LAMOST), Los Alamos National Laboratory, the Max-Planck-Institute for Astronomy (MPIA), the Max-Planck-Institute for Astrophysics (MPA), New Mexico State University, Ohio State University, University of Pittsburgh, University of Portsmouth, Princeton University, the United States Naval Observatory, and the University of Washington.\\

\section*{Data Availability}
After acceptance the code will be released in PaperData Repository powered by China-VO with a DOI number.\\


\bibliographystyle{mnras}
\bibliography{GPID} 





\bsp	
\label{lastpage}
\end{document}